\documentclass[useAMS,usenatbib]{mn2e}

\usepackage {graphicx}
\usepackage {times}
\usepackage{pdflscape}
\usepackage{rotating}
\usepackage{natbib}

\title[Spectroscopy of the specially-extended  Ly$\alpha$ emission around a QSO at z=6.4]{Spectroscopy of the spatially-extended Ly$\alpha$ emission around a QSO at z=6.4\footnotemark[0]
}
\author[Goto]{Tomotsugu Goto$^{1}$ %\footnotemark[1]
,
 Yousuke Utsumi$^{2}$,
 Jeremy R. Walsh$^{3}$,
Takashi Hattori$^{4}$,
\newauthor 
Satoshi Miyazaki$^{2}$,
and Chisato Yamauchi$^{5}$
\\
$^{1}$Institute for Astronomy, University of Hawaii
2680 Woodlawn Drive, Honolulu, HI, 96822, USA, tomo@ifa.hawaii.edu\\
$^{2}$Department of Astronomical Science,The Graduate University for Advanced Studies, 2-21-1 Osawa, Mitaka, Tokyo 181-8588, Japan\\
$^3$European Southern Observatory, Karl-Schwarzschild Strasse 2, D-85748, 
Garching, Germany\\\
$^{4}$Subaru Telescope 650 North A'ohoku Place Hilo, HI 96720, USA\\
$^{5}$Institute of Space and Astronautical Science, Japan Aerospace Exploration Agency, 	     Sagamihara, Kanagawa 252-5210
}

\begin{document}
%--- DRAFTCOPY: COMMENT OUT IF NOT NEEDED -------
% Prints a large "DRAFT" diagonally across each page
% Does not show up in TeXview
% \typeout{Prints "DRAFT" on each page; does not show in TeXView}
% \special{!userdict begin /bop-hook{gsave 200 30 translate
% 65 rotate /Times-Roman findfont 216 scalefont setfont
% 0 0 moveto 0.90 setgray (DRAFT) show grestore}def end}
%%------------------------------------------------
\def\Hg{H$\gamma$}
\def\Hd{H$\delta$}

%\date{Accepted 2006 December 15. Received 1988 December 14; in original form 2006 March 17}
\date{\today; in original form 2011 August 9}

\pagerange{\pageref{firstpage}--\pageref{lastpage}} \pubyear{2009}

\maketitle

\label{firstpage}

\begin{abstract}
We have taken a deep, moderate-resolution Keck/Deimos spectra of QSO, CFHQS2329, at z=6.4.
At the wavelength of Ly$\alpha$, the spectrum shows a spatially-extended component, which is significantly more extended than a stellar spectrum, and also a continuum part of the spectrum. The restframe line width of the extended component is 21$\pm$7 (\AA), and thus smaller than that of QSO (52$\pm$4\AA)
, where they should be identical if the light is incomplete subtraction of the QSO component.
Therefore, these comparisons argue for the detection of a spatially extended Ly$\alpha$ nebulae around this QSO.
 This is the first z$>$6 QSO that an extended Ly$\alpha$ halo has been observed around.
 Careful subtraction of the central QSO spectrum reveals a lower limit to the $Ly\alpha$ luminosity of  (1.7$\pm$0.1)$\times10^{43}$ erg s$^{-1}$.
 This emission may be from the theoretically predicted infalling gas in the process of forming a primordial galaxy 
that is ionized by a central QSO. 
On the other hand, if it is photoionized by the host galaxy, an estimated star-formation rate of  $>$3.0 $M_{\odot}$ yr$^{-1}$ is required.

If we assume the gas is virialized, we obtain dynamical mass estimate of $M_{dyn}$=1.2 $\times$10$^{12} M_{\odot}$.
The derived $M_{BH}$/$M_{host}$ is 2.1 $\times 10^{-4}$, which is two
 orders smaller than those from more massive z$\sim$6 QSOs, and places
 this galaxy in accordance with the $local$ M-$\sigma$ relation, in contrast to a previous claim on the evolution of M-$\sigma$ relation at z$\sim$6. 
We do not claim evolution or non-evolution of the M-$\sigma$ relation based on a single object, but
our result highlights the importance of investigating fainter QSOs at z$\sim$6. 
%Extended Ly$\alpha$ emissions around a QSO provide us with a unique opportunity to investigate QSO hosts fainter than can currently be detected in molecular emission lines. 

% We investigate neutral hydrogen absorption in a deep, moderate-resolution Keck/Deimos spectra of a QSO, CFHQS2329, at z=6.4.
% This QSO is one of the highest redshift QSOs presently known at z=6.4. Since the luminosity of the QSO is by 2.5mag fainter than a previously well-studied QSO, SDSSJ1148, another at z=6.4, it allows us to probe hydrogen absorption closer to the QSO (emission) redshift, and thus, to higher (absorption) redshift.
% The average transmitted flux at $5.915<z_{abs}<6.365$ (200 comoving Mpc) is consistent with zero, in all \lya, \lyb and \lyg absorption measurements. 
% This corresponds to the lower limit of optical depth, $\tau_{eff}^{\alpha},>4.9$.
% This is a detection of a complete Gunn-Peterson trough at  $5.835<z_{abs}<6.365$ in the line of sight of this QSO, as was seen with a QSO at z=6.28 previously.
\end{abstract}


\section{Introduction}

\begin{table*}
% \centering
 \begin{minipage}{180mm}
  \caption{Target information adopted from \citet{2007AJ....134.2435W} and \citet{2009MNRAS.400..843G}. }\label{tab:targets}
  \begin{tabular}{@{}lcccccccc@{}}
  \hline
   Object &  z$_{MgII}$&              $i'_{AB}$ & $z'_{AB}$ & $z_{R}$ & $J$                                  &  \\ 
% \hline
 \hline
CFHQS J232908.28-030158.8  & 6.417$\pm$0.002 & 25.54$\pm$0.02 & 21.165$\pm$0.003&  21.683$\pm$0.007          & 21.56$\pm$0.25 \\
\hline
\end{tabular}
\end{minipage}
\end{table*}

 Understanding the first stage of the formation of galaxies is one of the central issues in observational astronomy.
 Galaxy formation models predict that an early stage inevitably involves a spatially extended distribution of infalling, cold gas \citep{2001ApJ...556...87H}. If such gas is ionized by a central luminous quasar (QSO), or an initial starburst, such gas should be observed as  extended Ly$\alpha$ emission in the high-z Universe.

However, due to instrumental limitations, extensive searches for 
extended Ly$\alpha$ emission or Ly$\alpha$ blobs were mostly conducted 
at at z$\sim$2-3 in the past \citep{2000ApJ...532..170S,2004AJ....128..569M,2006A&A...452L..23N,2007MNRAS.378L..49S,2009MNRAS.393..309S}, %finding $L_{Ly\alpha}$ of 10$^{42}$-10$^{44}$ erg s$^{-1}$. 
with only a few exceptions at z=6.595 %with  $L_{Ly\alpha}$=3.9$\pm$0.2 $\times$10$^{44}$ erg s$^{-1}$ 
\citep{2009ApJ...696.1164O} and at z=4.5 \citep{2003Ap&SS.284..357B}.
 It is worth noting that some of these detections are associated with a bright QSO \citep{2003Ap&SS.284..357B,2005A&A...436..825W,2008MNRAS.389..792B}.
 If a QSO is a heating source, it is also a subject of interest to
 understand what role a central black hole plays in such an early stage
 of the galaxy formation, perhaps leading to the tight correlations observed at low redshift such as the M-$\sigma$ relation \citep{1998AJ....115.2285M}, and the starburst-AGN connection \citep{1997ApJ...485..552M,2005MNRAS.360..322G,2006MNRAS.369.1765G}.
%starburst AGN connection here?

However, at z=3, the age of the Universe was already 2.2 Gyrs, and thus,
it may be too late to search for a primordial stage of a galaxy formation \citep{2003ApJ...589...35S}. 
% In addition, most of the above Ly$\alpha$ blobs were not associated with a luminous QSO, and thus, the connection between galaxy evolution and QSO formation was unclear. 
In \citet{2009MNRAS.400..843G}, we found a spatially-extended structure around a QSO at z=6.4. This may be the first example of $Ly\alpha$ blob around a luminous QSO at z$>$6.
However, the detection was in the $z'$-band image of Subaru/Suprime-Cam. 
Therefore, it was not clear if the extended structure was Ly$\alpha$ emission, or continuum emission from the QSO host galaxy.
In this work, we performed a deep, moderate-resolution spectroscopy with the Keck/Deimos to clearly separate the two cases.
  Unless otherwise stated, we adopt the WMAP cosmology: $(h,\Omega_m,\Omega_L) = (0.7,0.3,0.7)$.% \citep{2009ApJS..180..330K}. 

% It has been observationally known that there is a tight correlation between the bulge luminosity and the mass of the central black holes observed in nearby inactive elliptical galaxies (Ferrarese 2006).
%
%the correlation between the super massive black holes and galaxy evolution.

 %   \end{landscape}

%\clearpage

\section{Observation}
\label{Observation}

%RA : 23:29:08.28
%DEC : -03:01:58.8
%
%CFHQS J232908-030158...	23 29 08.28	-03 01 58.8	>25.08a	21.76 ¡Þ 0.05	21.56 ¡Þ 0.25	>3.32	0.18 ¡Þ 0.25
%
%CFHQS J2329-0301...	6.43	Gemini	2006 Nov 26	1300	1.0	3600	0.7	-25.23

Our target is QSO CFHQS J2329-0301\citep[Table \ref{tab:targets};][]{2007AJ....134.2435W} at z=6.417 \citep{2010AJ....140..546W}.
% This is one of the three highest redshift QSOs to date, along with SDSSJ1148+5251 at z=6.419 \citep{2003AJ....125.1649F}  and CFHQS0210-0456 at z=6.438 \citep{2010AJ....140..546W}. 
This QSO is known to be in a dense environment surrounded by 7 LBG candidates\citep{2010ApJ...721.1680U}.
%  and to have extended Ly$\alpha$ emission and continuum \citep{2009MNRAS.400..843G}.
%The characteristics of the target are summarized in Table \ref{tab:targets}. 

We obtained moderate resolution spectra of the QSO using Keck/Deimos on
the nights of September 12 and 13, in 2010.
The details of the observation and data reduction are described in \citet{2011MNRAS.415L...1G}.
% Although Deimos was used in the multi-slit mode to investigate the wider environment of the QSO, we focus on the QSO specra in this paper.
Briefly, we used the 830 lines mm$^{-1}$ grating and the OG550 order cut filter with the  central wavelength of 8500\AA~. 
The slit width was 1.0'' with 0.47 \AA~pixel$^{-1}$, giving a resolving power of $R\sim$3600.
 The position angle of the slit was -30 deg.
The wavelength coverage was 6000\AA\ to 10000\AA.
 The spatial resolution is 0.1185'' pixel$^{-1}$. 
The total integration time was 5.5 hours.
 
 We used the Deep2 pipeline to reduce the data, except the background subtraction, which we did manually to remove ghost features of the 830 grating carefully.
 Wavelength calibration is based on the HeNeAr lamp. 
%Fig. \ref{fig:2d} shows two-dimensional spectrum of the QSO. 

%After the standard reduction, there remained a jump in the continuum level between red and blue CCDs. Therefore we have adjusted the continuum level in the blue CCD to that of the red, which is flux calibrated. The CCD gap is at 8400\AA for the QSO spectrum.
 
%At $>$7500\AA, many strong sky OH emission lines are present. Although the DEEP2 pipeline subtract these sky lines well, the noise level at the exact wavelengths of the sky lines are significantly higher than the other wavelength clear of emission. Therefore, we have masked out these wavelengths of strong sky emissions. 
%The green lines at the bottom shows the sky spectrum in an arbitrary unit.

For flux calibration, the spectrophotometric standard G191-B2B was observed and used to correct the spectral shape.
Absolute flux calibration was achieved by passing the spectra of the QSO through the Subaru $z'$ filter and normalizing to match the observed $z'$ magnitude in Table \ref{tab:targets}.
We independently performed the absolute flux calibration using the standard G191-B2B, obtaining only 4\% larger absolute flux.

%\subsection{Revisiting Redshift Determination}

%
%
%\begin{figure*}
%\begin{center}
%\includegraphics[scale=0.9]{/home/tomo/paper/GPtest_J2329/GPtest_J2329_v1/2011-02-02_J2329_deimos_1dlong_s.ps}
%\end{center}
%\caption{
%One-dimensional spectra of CFHQSJ2329, but showing wider wavelength range.%, boxcar smoothed with 20pix.
%The lines are the same as in Fig.\ref{fig:1d}.
%}\label{fig:1dlong}
%\end{figure*}
%

\begin{figure}
\begin{center}
\includegraphics[scale=0.55]{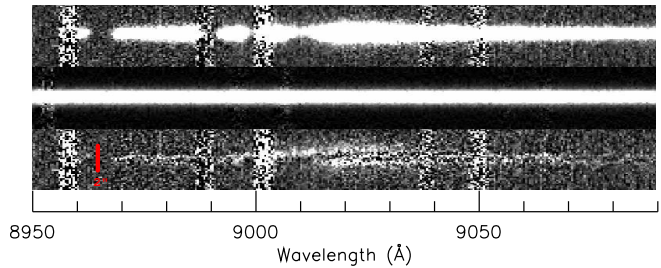}
\end{center}
\caption{
Subtracting stellar spectrum from QSO spectrum.
The top panel shows the QSO spectrum. The middle panel is a reference PSF stellar spectrum.
%The botoom panel shows residual from subraction of the stellar spectrum from the QSO. 
The bottom panel shows the residuals from the subtraction of the 
PSF spectrum and the smooth extended component from the QSO spectrum.
Pixel scale in spatial direction is 0.1185'' pix$^{-1}$. The red bar shows a scale of 2''.
}\label{fig:2dsubtraction}
\end{figure}

\section{Analysis}\label{analysis}

\subsection{Removal of the QSO PSF}

\begin{table}
\begin{center}
\caption{
Properties of Ly$\alpha$ emission from the extended region. 
}
\label{tab:line_properties}
\begin{tabular}{llll}
  \hline
    FWHM (\AA) & Flux (erg cm$^{-2} s^{-1})$ & Luminosity (erg s$^{-1}$) & Velocity Disp. (km s$^{-1}$)\\ 
 \hline
   21$\pm$7  & (3.6$\pm$0.2) $\times10^{-17}$  &  (1.7$\pm$0.1)$\times10^{43}$ &301$\pm$99 \\
\hline
\end{tabular}
\end{center}
\end{table}

To investigate the extended component of the spectrum, we need to subtract the central point source spectrum, which is often brighter by a factor of $>$10 for a QSO. 
We chose a spectrum of a bright
star in the same Deimos mask as the QSO and which was relatively 
free of the ghost features of the 830G grism; this spectrum is shown
in the middle panel of Fig.\ref{fig:2dsubtraction}.
%We picked a spectrum of a bright star, which is relatively free from ghost features of the 830G grism, in the same deimos mask for this purpose (the middle panel of Fig.\ref{fig:2dsubtraction}). 
The PSF of the QSO and the extended background were decomposed (in the
spatial direction) using the IRAF code {\it specinholucy} 
based on a two-channel restoration technique \citep{2003AJ....125.2266L}. 
The technique restores the spatial profile of the PSF and a 
smooth background wavelength-by-wavelength.
%For fitting and subtraction, we used  the IRAF code {\ttfamily specinholucy } \citep{2003AJ....125.2266L}. 
It is based on a two-channel restoration algorithm
that restores a point spread function (PSF)-like component in a
2D spectrum and an underlying extended background component.
It has already been successfully used to subtract point-source spectra in highly
inhomogeneous backgrounds, such as high-z SNe Ia embedded in their host
galaxy \citep{2005A&A...431..757B}. Therefore, it is also suitable for
our case of separating QSO from the extended structure.

 The top panel of Fig.\ref{fig:2dsubtraction} shows the reduced 2D spectrum of J2329.
 The middle panel shows the spectrum of the reference PSF star.
 The PSF size (FWHM) measured from a star in a slit is 0.65''.

% The bottom panel is the residual from the PSF spectrum subtraction.
% The fit and subtraction was performed wavelength by wavelength.

The restored background component was smoothed by a Gaussian  \citep[see ][ for details]{2003AJ....125.2266L} of FWHM 16.5 pixels (1.95''). On 
subtracting this smooth component from the spectrum (with the QSO PSF 
removed), structure at spatial scales intermediate between that of the 
PSF and the smooth extended structure was revealed. The bottom panel 
of Fig.\ref{fig:2dsubtraction} shows this revealed component.

In the bottom panel of Fig.\ref{fig:2dsubtraction}, at the peak of the QSO spectrum,
there remain positive and negative residuals resulting from
incomplete PSF removal. This has often been seen in PSF removal 
and usually arises when the spatial profile of the PSF star does not 
quite match that of the target. The residuals are at the level of a 
few percent of the peak of the QSO PSF. 
% In the bottom panel of Fig.\ref{fig:2dsubtraction}, at the center of the spectrum, there remains residual of incomplete PSF subtraction in a horizontal direction. This has often been seen in PSF subtractions because of the fixed pixel size, the S/N of the fit is poorer at the center of the PSF, where counts are much higher. 

\begin{figure}
\begin{center}
\includegraphics[scale=0.45]{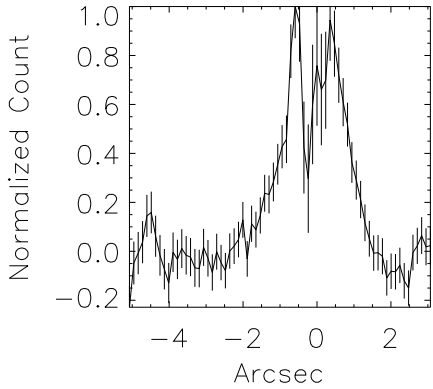}
\end{center}
\caption{
The residuals on the two channel restoration of the QSO 2D spectrum
are shown summed in wavelength from 9006 to 9035\AA. The error bars
include those of
the QSO, the star, and the background noise.
}\label{fig:hist}
\end{figure}

Noteworthy in Fig.\ref{fig:2dsubtraction} is the extended flux over the wavelength range 
9006-9035\AA\ that remains from the decomposition. The peak of this
flux corresponds to Ly$\alpha$ wavelength at the redshift of the QSO.
The red bar of length 2'' in the figure shows the spatial scale for
reference, indicating that the remaining flux is spatially extended 
over an extent as large as $\sim$4''. In Fig.\ref{fig:hist}, we show the residual extended 
flux from the two-channel restoration summed in the wavelength 
direction over the extent 9006 to 9035\AA. The error bars include 
Poisson noise from the QSO and star spectra, and 
%a Monte Carlo generated Gaussian noise estimate from the restoration. 
the background noise.
Compared with the
errorbars, remaining flux from -1.7 to +1.6 arcsec appears
to be significant.

% Apart from this, at around 9006-9035\AA~ there remains extended flux. At this redshift, this corresponds to Ly$\alpha$ wavelength at 9017\AA.
% The pixel scale in the spatial direction is  0.1185'' pixel$^{-1}$. We put a bar of 2'' length in the figure for a reference, showing  that the remaining flux are spatially extended and as large as 2''.
 
% To assess the accuracy of the fit, we performed an independent PSF fit and subtraction after summing the data from 9018 to 9025\AA.
% Both QSO and fitted stellar PSF are shown in Fig.\ref{fig:cross} in the black and red lines. Residuals of the subraction of these are shown in the green line, which agrees well with that from the fit by {\ttfamily specinholucy}.

%\begin{figure}
%\begin{center}
%\includegraphics[scale=0.34,angle=270]{/mnt/4.5TB/data/keck/2010sep13_analysis_v3/from_jeremy/Fig2.ps}
%\end{center}
%\caption{
%Cross section at the Lyman$\alpha$ line (X=145-159).
%The results from the two-dimentional spectra subtraction is in the blue line.
%The QSO and stellar spectra are black and red, respectively.
%The green line shows the residual of the subraction after summing X=145-159.
%}\label{fig:cross}
%\end{figure}

%
%In Fig.\ref{fig:hist}, we show residuals summed in the wavelength direction from 9006 to 9035\AA. 
%Error bars include poisson noise from QSO and star spectra, and background noise. 
%Compared with the errorbars, remaining flux from -2 to +2 arcsec are significant.

\begin{figure}
\begin{center}
\includegraphics[scale=0.45]{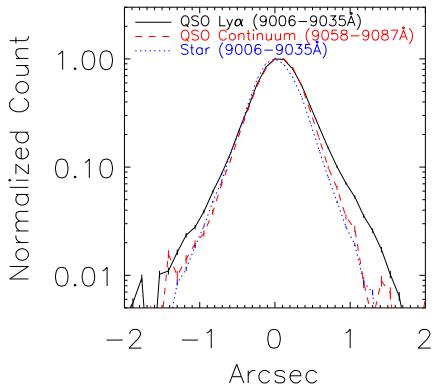}
\end{center}
\caption{
Comparison of the spatial profiles of the QSO spectrum over the
Ly-$\alpha$ line
between 9006 and 9035\AA\ (black solid line), and over the continuum
between 9058
and 9087\AA\ (red dotted line). The blue dotted line show the spatial
profile of a star
between 9006 and 9035\AA.
%Comparing spatial profiles of the QSO spectrum between 9006 and 9035\AA (in the black solid line), and between 9058 and 9087\AA (red dotted line). The blue dotted line show the spatial profile of a star between 9006 and 9035\AA.
}\label{fig:hist_conti}
\end{figure}

To further verify this finding, in Fig.\ref{fig:hist_conti}, we compare the spatial flux profile in the wavelength range between 9006 and 9035\AA\  of the QSO spectrum (corresponding to Ly$\alpha$ emission) in the black solid line,
 and that between 9058 and 9087\AA  (i.e., remote from the Ly$\alpha$ line) in the red dotted line. 
 Both of these wavelength ranges are chosen so that they do not include strong sky emission lines. 
 The flux profile of the star in the range of 9006-9035\AA\ is also plotted in the blue dotted line. 
Compared with the associated errorbars, the flux profile of the QSO at the  Ly$\alpha$ wavelength is significantly more extended than that of the continuum, and the star.
We also performed a 2-dimensional subtraction using the QSO continuum as a spectral PSF, interpolating the PSF shape from the continuum on both sides of the Ly$\alpha$ line, obtaining a very similar result as in Fig.\ref{fig:2dsubtraction}.
Note that the continuum region may contain an extended component from the underlying galaxy.
%The comparison between the star and the QSO spectra was subject to a possible PSF variation due to slightly different optics.
%However, this comparison is robust to such variation since the light in the same slit are through the same optics.
%According to the Kolmogorov-Smirnov test, these two distributions are different at xxxx \% significance.
% The comparison between the PSF of the star and the QSO spectrum 
%is subject to possible PSF variation as the two spectra are taken 
%from different regions of the field where 
%The optical distortions may differ between the star and QSO slit positions. 
%However, the spatial profiles of the star and QSO continuum are very similar, suggesting such effects are small.
 This test was performed in case the optical distortions in the
spectrograph differ between the position of the star used as a PSF and QSO slit position.
However, the spatial profiles of the star and QSO continuum are very similar (see
Fig. \ref{fig:hist_conti}), suggesting such effects are small.

We note that the extended emission may be stronger above than below the QSO
spectrum in Fig.\ref{fig:2dsubtraction}. Corresponding asymmetry can be found in Fig.\ref{fig:hist}. This
can also be seen in Fig.\ref{fig:hist_conti}, where the excess from the continuum subtraction is larger
at the positive offset. 
Also the emission may be tilted in wavelength-position space
as revealed in the bottom panel of Fig.\ref{fig:2dsubtraction}. This may be a sign of rotation
in the extended Ly$\alpha$ emitting region, but needs confirmation from a better spatial
resolution spectrum.

% We would like to mention that the extended emission may be stronger on the upper side in Fig.\ref{fig:2dsubtraction}. Corresponding asymmetry can be found in Fig.\ref{fig:hist}. This can also be seen in Fig.\ref{fig:hist_conti}, where the excess from the continuum is lager at the positive offset. 
%Also the emission may be tiled in the bottom panel of Fig.\ref{fig:2dsubtraction}. This may be a sign of rotation, but needs confirmation from a better spatial resolution spectrum.

\subsection{One dimensional spectrum}

On the basis of this evidence, we claim that at least the Ly$\alpha$ 
region of the spectrum is more extended than the PSF from the star 
and the continuum region of the QSO spectrum. Next, we extracted a one-dimensional spectrum from the 2D image by masking out the central 10
pixels (1.1''), where the residuals on the PSF match are strong.
Note that due to this masking, the computed flux and luminosity are 
lower limits.

%With these evidences, we claim that at least Ly$\alpha$ region of the spectrum are more extended than the PSF from the star and the continuum region of the same spectrum. 
%Next, we extract one-dimensional spectrum from 2D image by masking out central 10 pixels (1.1''), where residual are affected by the low S/N of the fit at the center. Note that due to this masking, the computed flux and luminosity are lower limit. 

\begin{figure}
\begin{center}
\includegraphics[scale=0.6]{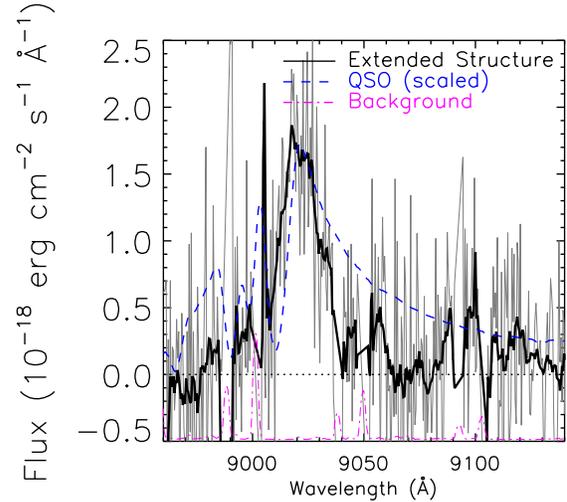}
\end{center}
\caption{
One-dimensional spectra of the extended region of CFHQSJ2329 after subtracting the central QSO.%, boxcar smoothed with 20pix.
The gray line shows spectrum in the original resolution.
The black line shows binned spectrum with 10 pixel box.
The magenta dash-dotted lines show the sky spectrum in an arbitrary unit. 
%The corresponding two-dimensional spectrum is shown in the bottom panel of Fig.\ref{fig:2dsubtraction}.
The blue dotted line shows the arbitrary-scaled spectrum of the QSO.
%, whose Ly$\alpha$ emission is significantly broader than that of the extended region.
}\label{fig:1d}
\end{figure}

In Fig.\ref{fig:1d}, we show the one-dimensional spectrum. 
%The gray lines are in the original resolution. 
The gray line shows the spectrum at the original resolution. The black line
shows the same spectrum binned with a 10 pixel box. The magenta  line shows
the scaled sky spectrum. 
%The black solid line shows a spectrum binned with a 10 pixel box. 
As expected from Fig.\ref{fig:2dsubtraction}, a strong peak can be seen at around Ly$\alpha$ emission.
 Due to neutral hydrogen absorption on the bluer side,
the line has clear asymmetric profile with an extended red wing that is typical for a high-z Ly$\alpha$ line.
In the blue dotted line, we show the arbitrary-scaled QSO spectrum before the PSF subtraction.
The Ly$\alpha$ line profile of the QSO is much broader. % The Ly$\alpha$ line profile is very different from that of the residual.
This difference confirms that the residual in the bottom panel of Fig.\ref{fig:2dsubtraction} is not part of the (unsubtracted) QSO spectrum, but indeed spectrum of a different nature.

The total flux measured between 9006 and 9035\AA~region (except central 1'') is 
(3.6$\pm$0.2) $\times10^{-17}$ erg cm$^{-2} s^{-1}$ as shown in Table \ref{tab:line_properties}. Compared with errors measured from 9058-9087\AA~region of the spectrum, 
this is a 17 $\sigma$ detection. There is no significant flux detected other than the Ly$\alpha$ wavelength.
 The flux corresponds to the lower limit of Ly$\alpha$ luminosity of (1.7$\pm$0.1)$\times10^{43}$ erg s$^{-1}$
(c.f., the  Ly$\alpha$ luminosity of the QSO is 6.2$\times10^{44}$ erg s$^{-1}$).
This is smaller than photometrically estimated value of 2.0$\times10^{43}$ erg s$^{-1}$\footnote{\citet{2009MNRAS.400..843G} had a computational error. We quoted a corrected value here.} \citep{2009MNRAS.400..843G}, but consistent with each other, since our estimate is a lower limit due to the central mask.
For comparison, the luminosity of a Ly$\alpha$ blob at $z\sim$6.5 is $L^*\sim$3.9$\pm0.2\times$10$^{43}$ erg s$^{-1}$ \citep{2009ApJ...696.1164O} and Ly$\alpha$ blobs at z=3.1 is $\sim$1$\times$10$^{43}$ erg s$^{-1}$ \citep{2004AJ....128..569M,2006ApJ...640L.123M}. 
%For comparison, the median luminosity of Ly$\alpha$ blobs at z=6.6 is 110 km s$^{-1}$ \citep{2009ApJ...696.1164O}, and at z=3.1 is 330 km s$^{-1}$ \citep{2004AJ....128..569M,2006ApJ...640L.123M}.
%\cite{2011MNRAS.410L..13M} surveyed a large volume (1.6$\times$10$^{6} Mpc^3$) to find several LABs with several 10$^{43}$ erg s$^{-1}$.
% xxxxx numbers xxxxx.
% The luminosity could be larger than the brightest Ly$\alpha$ emitter (3.9$\pm$0.2 $\times10^{43}$ erg s$^{-1}$) from the entire 1 deg$^2$ of SXDS field \citep{2009ApJ...696.1164O}. 

 The FWHM estimated through a Gaussian fit is 21$\pm$7 (\AA) in the restframe. 
 This narrow width of the line also suggests that the Ly$\alpha$ emission did not originate from the broad line regions of the QSO (FWHM of 52$\pm$4\AA).
 We do not recognize the presence of a clear rotation curve, or multiple velocity components.
 In terms of velocity dispersion in the restframe, this corresponds to 301$\pm$99 km s$^{-1}$. 
%If the emission is gravitationally associated, the presence of a massive object such as a host galaxy is suggested. 
For comparison, the median velocity dispersion of Ly$\alpha$ blobs at z=6.6 is 110 km s$^{-1}$ \citep{2009ApJ...696.1164O}, and at z=3.1 is 330 km s$^{-1}$ \citep{2004AJ....128..569M,2006ApJ...640L.123M}.
%For comparison, the median velocity dispersion of Ly$\alpha$ emitters at z=6.6 is $\sim$130 km s$^{-1}$ \citep{2005PASJ...57..165T}, and at z=3.1 is  110 km s$^{-1}$ \citep{2005A&A...431..793V}.
%Therefore, it is most likely that this emission is from gravitationally-bounded nebulae, heated by a QSO or a star-formation.

 If the nebula forms a single virialized system with a velocity dispersion of 301 km s$^{-1}$ in 2'' radius (11 kpc at z=6.4), we estimate a virial mass of 1.2 $\times$10$^{12} M_{\odot}$ (5/3$\times$3$\sigma^2 R/G$).
 For comparison, \citet{2009MNRAS.400..843G} estimated a stellar mass from 6.2$\times 10^8$ to 1.1$\times 10^{10} M_{\odot}$ depending on the star-formation history.
If the velocity dispersion of the nebula reflects the dynamics of the host galaxy, it suggests that a massive galaxy is already harboring a QSO at z=6.4. 
%We discuss the heating source of the nebula in the next section. 

%velocity width is dv(km/s)=c*dlambda/lambda
%3E5*20/9030=664. 

\begin{figure}
\begin{center}
\includegraphics[scale=0.5]{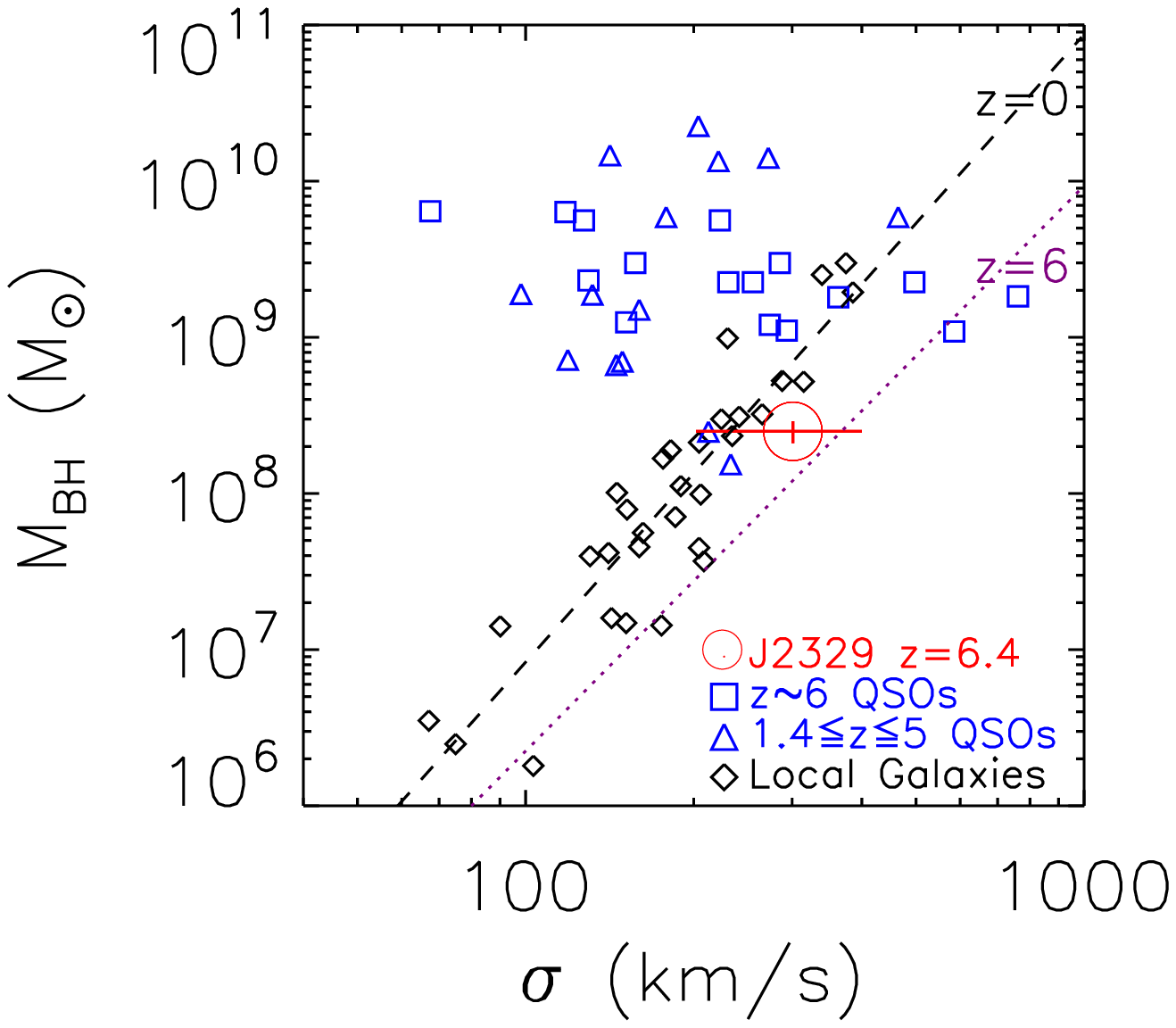}
\end{center}
\caption{
$M_{BH}-\sigma$ relation. 
The data points except for J2329 are adopted from \citet{2010ApJ...714..699W}.
%The black hole masses of the z$\sim$6 quasars 
%versus their bulge velocity dispersion ($\sigma$). 
%$\rm log\,(M_{BH}/M_{\odot})=8.13+4.02log(\sigma/200\,km\,s^{-1})$ \citep{2002ApJ...574..740T}.
The diamonds represent the local galaxies \citep{2002ApJ...574..740T}, whose local $\rm M_{BH}-\sigma$ relationship in the black dashed line is $\rm log\,(M_{BH}/M_{\odot})=8.13+4.02log(\sigma/200\,km\,s^{-1})$.
The blue squares and triangles
are for the z$\sim$6 and $\rm 1.4\leq z\leq5$ QSO samples, 
respectively, with $\sigma$ derived from the observed CO 
line width using $\sigma\approx FWHM/2.35$  \citep{2010ApJ...714..699W}. 
The red circle show J2329 at z=6.4 measured in this work based on the dispersion of the Ly$\alpha$ line.
The purple dotted line shows a simulated M-$\sigma$ relation at z=6 \citep{2006ApJ...641...90R}.
}\label{fig:Msigma}
\end{figure}

\subsection{Non-detection of Continuum}

\citet{2009MNRAS.400..843G} detected 2.5$\times$10$^{-19}$~erg~cm$^{-2}$~s$^{-1}$~\AA$^{-1}$ of flux density from the extended component in $z_r$ filter ($\lambda_e$=988nm) at 3 $\sigma$. We did not detect any continuum from the extended component other than the Ly$\alpha$ emission. However, the noise level of our continuum is 5.5$\times$10$^{-19}$~erg~cm$^{-2}$~s$^{-1}$~\AA$^{-1}$ pixel$^{-1}$. 
Therefore, we cannot indicate the detection or non-detection of the possible continuum emission from the extended component.

Since we did not detect continuum from the extended component, we cannot
obtain an equivalent width (EW) solely from the spectra.
However, if we use the continuum level of  $z_r$ band, we obtain the restframe EW of 19$\pm$6 \AA.~
This is not a particularly large value for Ly$\alpha$ blobs. 
For example, all 18 Ly$\alpha$ blobs in \citet{2008ApJ...675.1076S} had EW$_{rest}>100$\AA~(but not all of them are associated with a QSO).

\section{Discussion}\label{discussion}

\subsection{Comparison to  Ly$\alpha$ blobs around low-z QSOs}
It is informative to compare the size of the extended Ly-$\alpha$
emission around CFHQS J232908.28-030158.8 with that of similar structures, Ly-$\alpha$ blobs around lower redshift QSOs.
\citet{2008MNRAS.389..792B}'s Ly$\alpha$ nebula around a z=2.48 QSO had  FWHM$_{rest}$ of 370$\pm$63 km/s, with 40 kpc radius of extension (in 3 separate parts though).
The Ly$\alpha$ halo around a radio-quiet QSO at z=3.04 also extended to $\sim$30 kpc \citep{2005A&A...436..825W}.
\citet{2003Ap&SS.284..357B} reported a 5''-extended, FWHM$_{rest}$ of 181 km/s Ly$\alpha$ nebula around a QSO at z=4.5.
Compared to these, our z=6.4 Ly$\alpha$ nebula has a much larger FWHM$_{rest}$ of 707$\pm$232 km/s, perhaps reflecting a larger $M_{host}$. Although the 11 kpc of extension is smaller than the measured sizes at lower-z, surface brightness dimming is much greater at z=6.4, and thus, the size could be larger than measured. 
There exist only a few examples of extended Ly$\alpha$ halos around QSOs. It is important to construct a larger sample to correctly understand the evolution of the Ly$\alpha$ halo around QSOs.

\subsection{What is the physical origin of this Ly$\alpha$ nebula?}
%What is the physical origin of this Ly$\alpha$ nebula?
Since there exists a QSO at the center of this nebula, an obvious explanation is halo gas photoionized by the QSO.
 Following \citet{2005ApJ...620...31Y}, we estimate the ionizing photon emission rate of this QSO using absolute magnitude $M_{1450\AA}$=-25.23 \citep{2007AJ....134.2435W}. We count photons with energy in the range 13.6-54.4 eV as ionizing photons.
We assumed average QSO SEDs at $z>3$ from \citet{2002ApJ...565..773T}. To be conservative, we used the SED of radio-loud QSOs, which produce smaller ionizing photon rates than radio-quiet QSOs. The rate obtained for the QSO is 1.6$\times$10$^{56} s^{-1}$.
On the other hand, Ly$\alpha$ photon rate emitted by the nebula is several
  $ 10^{53} s^{-1}$. 
Therefore, there are orders more ionizing photons produced by the QSO to illuminate the Ly$\alpha$ nebula.
A cold accretion of neutral gas \citep{2003Natur.421..341B}, or recently found re-scatted Ly$\alpha$ photons by neutral hydrogen \citep{2011Natur.476..304H} are other candidates to explain Ly$\alpha$ blobs. In our case, however, these scenarios are unlikely because neutral hydrogen cannot survive exposure to the abundant ionizing photons.
% We detected the extended Ly$\alpha$ emission around the QSO.  
% This may be theoretically predicted infalling gas forming a galaxy, illuminated by a central QSO \citep{1988MNRAS.231P..91R}. If so, quantifying this will provide us with an important constraint on the formation of galaxies such as gas fraction, AGN geometry, and covering factors in the early Universe \citep{2001ApJ...556...87H}.
% 

In addition, star-formation in the host galaxy can also contribute \citep{2000ApJ...532L..13T,2003ApJ...591L...9O}.
The star-formation rate (SFR) can be estimated using the following relation \citep{1998ARA&A..36..189K,2007PASJ...59..277T}.
\begin{equation} \label{eqn:equation}
 SFR(Ly\alpha) = 9.1 \times 10^{-43} L(Ly\alpha) M_{\odot} yr^{-1},
\end{equation}
 Based on the Ly$\alpha$ luminosity, we estimate the SFR(Ly$\alpha$) is  $>$3.0 $M_{\odot}$ yr$^{-1}$. 
This is not a high SFR, and thus, can be expected from a young star-forming galaxy.
Therefore, it is possible that the star-formation in the host galaxy contributes to ionizing the Ly$\alpha$ nebula.
% However, the SFR estimate here is a lower limit because a significant fraction of bluer side of the Ly$\alpha$ emission could be absorbed by the neutral hydrogen. The flux from the central part overlapping with QSO is not taken into account.

%Another possibility is cold accretion of neutral, filamentary gas, producing Ly$\alpha$ emission through collisional excitation \citep{2006A&A...452L..23N,2010MNRAS.407..613G,2010ApJ...725..633F}. This scenario might also contribute since this QSO is in a 7 times over-dense region at the center of proto-cluster \citep{2010ApJ...721.1680U}. 
%Not all QSOs at z$\sim$6 are in a dense environment \citep{2009ApJ...695..809K}. It may not be a coincidence that the first z$\sim$6 QSO with a Ly$\alpha$ halo exists in a dense environment.
%
%
%
%Recently, polarized Ly$\alpha$ emission has been observed from Ly$\alpha$ blobs, 
%showing these Ly$\alpha$ photons were not produced in the nebula but in the galaxy to be re-scatted by neutral hydrogen \citep{2011Natur.476..304H,2011ApJ...736..160S}. 

\subsection{$M_{BH}-\sigma$ relation at $z>$6}
The virial mass estimate provides an interesting opportunity to
investigate the $M_{BH}$/$M_{host}$ ratio at z$>6$.
The black hole mass of this QSO was measured to be 2.5$\pm{0.4}\times10^8 M_{\odot}$ based on the MgII line and $L_{3000\AA}$ \citep{2010AJ....140..546W}. Combined with our virial mass of  1.2 $\times$10$^{12} M_{\odot}$, we obtain $M_{BH}$/$M_{host}$ of 2.1 $\times 10^{-4}$. This value is much smaller than those previously measured from brighter QSOs at z$\sim$6. 
For example, \citet{2010ApJ...714..699W} obtained a median  $M_{BH}$/$M_{bulge}$ ratio of 0.022 for bright ($M_{BH}>10^9 M_{\odot}$) QSOs at z$\sim$6, leading to a discussion that super massive BHs at z$\sim$6 grow rapidly without commensurate growth of their host galaxies.
On the contrary, this QSO is consistent with the local M-$\sigma$ relation as shown in Fig.\ref{fig:Msigma}.

Based on their galaxy merger simulations, \citet{2006ApJ...641...90R} predicted a weak redshift-dependent shift in the M-$\sigma$ relation due to an increasing velocity dispersion for a given galactic stellar mass. 
This QSO is even consistent with their predicted  M-$\sigma$ relation at z=6 shown in the purple dotted line in Fig.\ref{fig:Msigma}.
%Note, however, that our measurement is under assumption that the gas is virialized with no significant effect from resonant scattering.
Note, however, that our measurement is made under the assumption that 
the gas is virialized. It is assumed that there is no effect on 
the measured line profile from non-kinematic broadening mechanisms  
(such as resonant atomic scattering or dust scattering).

We do not claim an evolution or non-evolution of the $M_{BH}$/$M_{bulge}$ ratio based on one data point at z=6.4. 
However, due to the optical selection limit, previous QSO samples at  z$\sim$6 were limited to extremely massive BHs with $M_{BH}>10^9 M_{\odot}$, sampling only a small range in BH mass. Combined with errors in measuring velocity dispersion, this may have been one of the reasons why an almost flat $M_{BH}-\sigma$ relation was found at  z$\sim$6 \citep{2006ApJ...641..683S,2010ApJ...714..699W}, and shown in Fig.\ref{fig:Msigma}. Our result with a fainter QSO highlights the importance of sampling a wider range in $M_{BH}$ to more accurately assess the $M_{BH}-\sigma$ relation at z$\sim$6. Extended Ly$\alpha$ emission around a QSO may offer an alternative way to investigate fainter QSOs, whose molecular gas lines cannot be easily observed with existing facilities.

%%Consistent with zr?
%Consistent with z?
%M-sigma relation? MBH=2.5 e8 Msun
%
%implication for highz galaxy formation.
%lya En = -13.6 eV/n2
%
%\subsection{Physical origins of Ly$\alpha$ emission}
%QSO?
%Star formation?
%

\subsection{Note added at revision}

After this paper was submitted and a referee's report received, an
observational spectroscopic study of the identical QSO appeared on arXiv.
\citet{Willott2011} observed the same Ly-$\alpha$ halo with Keck/ESI
for 8.5 hours in seeing of 0.93'' They found L$_{Ly\alpha} >8 \times 10^{43}$
erg/s, with an extension of over 15 kpc, and a FWHM of 640 km/s. However,
in contrast to the work presented here, an independent PSF was not
available to remove the underlying QSO continuum. We show here that the
results are very similar if a PSF star or the QSO continuum are utilised
to remove the QSO across the Ly-$\alpha$ line, thus independently
strengthening the result in Willott et al. In consequence the numbers
for the extended Ly-$\alpha$ flux and width reported in Willott are
consistent with the findings in this paper.

%After this paper is submitted, a parallel work appeared on the arxiv.
%\citet{Willott2011} observed the same Ly$\alpha$ halo with Keck/ESI for 8.5 hours under the seeing of 0.93'' %(c.f. our exposure was 5.5 hours with the seeing of 0.65''). 
%They found $L_{Ly\alpha}>$8$\times 10^{43}$ erg/s, with extension of over 15 kpc, and FWHM of 640 km/s.
%Reassuringly these numbers are consistent with our findings in this paper.

%\section{Summary}

%\section*{Acknowledgments}

We thank the referee and M.Koss for insightful comments.

\bibliography{tomo_qso_v3} 
\bibliographystyle{mnras} 

%\appendix

%\section[]{PSF subtraction test}
%\label{sec:PSF-PSF}

%\bsp

\label{lastpage}

\end{document}